# Emergent togetherness in collaborative dance improvisation: neural and motor synchronization reveal a coupling-decoupling paradox

Yago Emanoel Ramos[1,2], Raphael Silva do Rosário[1], Adriana de Faria Gehres[3], Maria João Alves[4,5], Ana Maria Leitão[4,5], Cecília Bastos da Costa Accioly[6], Fatima Wachowicz[6], Ivani Lúcia Oliveira de Santana[7], José Garcia Vivas Miranda[1*]

**Abstract**

Collective improvisation in dance provides a rich natural laboratory for studying emergent coordination in coupled neuro-motor systems. Here, we investigate how training shapes spontaneous synchronization patterns in both movement dynamics and brain signals during collaborative performance. Using a dual-recording protocol integrating 3D motion capture and hyperscanning EEG, participants engaged in free, interaction-driven, and rule-based improvisation before and after a program of generative dance, grounded in cellular-automata. Motor behavior was modeled through a time-resolved α-exponent derived from Movement Element Decomposition scaling between mean velocity and displacement, revealing fluctuations in energetic strategies and degrees of freedom. Synchronization events were quantified using Motif Synchronization (biomechanical data) and multilayer Time-Varying Graphs (neural data), enabling the detection of nontrivial lead-lag dependencies beyond zero-lag entrainment. Results indicate that training produced an intriguing dissociation: inter-brain synchronization increased, particularly within the frontal lobe, while interpersonal motor synchrony decreased. This opposite trend suggests that enhanced participatory sense-making fosters neural alignment while simultaneously expanding individual motor explorations, thereby reducing coupling in movement. Our findings position collaborative improvisation as a complex dynamical regime in which *togetherness* emerges not from identical motor outputs but from shared neural intentionality distributed across multilayer interaction networks, exemplifying the coupling-decoupling paradox, whereby increasing inter-brain synchrony supports the exploration of broader and mutually divergent motor trajectories. These results highlight the nonlinear nature of social coordination, offering new avenues for modeling creative joint action in human systems.

**Keywords:** Motif-synchronization, Time-varying graphs, Movement element decomposition, Togetherness, Brain functional networks, Generative dance.

[1] Physics Institute, Federal University of Bahia, Salvador, Brazil.
[2] Physics Department, Federal University of Pernambuco, Recife, Brazil.
[3] Superior School of Physical Education, University of Pernambuco, Recife, Brazil.
[4] Faculty of Human Kinetics, University of Lisbon, Lisbon, Portugal.
[5] Institute of Ethnomusicology - Centre for Studies in Music and Dance, Lisbon, Portugal.
[6] Dance School, Federal University of Bahia, Salvador, Brazil.
[7] Department of Performing Arts, Federal University of Rio de Janeiro, Rio de Janeiro, Brazil.
[*] vivas@ufba.br

1. **Introduction**

   Movement synchrony refers to the temporal and spatial alignment of performers, a process tightly linked to social perception (intuition of social relationships) and aesthetic appreciation in dance (Cross et al., 2024). Studies show that moving in synchrony increases feelings of affiliation (Hove & Risen, 2009), collective enjoyment Vicary et al., 2017), and predicts pro-social effects (von Zimmermann et al., 2018). According to Orgs et al. (2024), in dance observation (within a video/fMRI context), higher degrees of movement synchrony between performers are associated with greater neural synchrony among spectators, which is interpreted as an indicator of shared aesthetic experience. Group-based coordination also shapes dancers' internal experience: distributed patterns of movement coupling enhance feelings of liking and group affiliation among performers (von Zimmermann et al., 2018). Neuroscientifically, these phenomena engage the action-observation network, mirror-neuron systems, and predictive motor pathways. As reviewed by Ribeiro et al. (2024), observing dance engages visual processing streams (dorsal and ventral visual pathways) that project to parietal and premotor cortices, activating the action observation network. These neural mechanisms support motor resonance and embodied processing during dance observation, potentially facilitating interpersonal coordination and contributing to the social and aesthetic impact of synchronized movement.

   Physiological evidence further demonstrates the embodied nature of synchronization (Noy et al., 2015). In improvisational tasks such as the mirror game, there exist moments of *togetherness*, which by Gesbert et al. (2022) define as the experience of being and acting together, characterized by tight kinematic alignment and subjective reports of shared flow. It fits within what De Jaegher & Di Paolo (2007) describe as *participatory sense-making*. Noy et al. (2015) showed that synchronized motion elicits shared physiological arousal (e.g., heart rate and respiration), reinforcing collective engagement and flow during reciprocal interaction. An fMRI hyperscanning study of dancers found distinct patterns of social network activation in leaders vs. followers, although it did not directly measure brain-to-brain synchrony (Basso et al., 2021; Chauvigné & Brown, 2018; Chauvigné et al., 2018). Taken together, these findings suggest that synchronization goes beyond movement alone. Moments of *togetherness* often emerge during synchronized creative processes, such as the improvisational mirror game, in which performers experience a sense of unity and flow through tightly coupled movements (Gesbert et al., 2022). These findings situate group coordination as not only a physical phenomenon but also an embodied, interdependent social process.

   In addition, within group-based coordination, leadership dynamics often influence the degree of synchrony achieved: leaders tend to guide timing and complexity of movement, while followers exhibit predictive and reactive adjustments (Basso et al., 2021). This role differentiation is critical, as it balances flexibility with structured coordination. In dance improvisation, for instance, shifts between leader and follower create opportunities for creativity while maintaining collective engagement. While significant advances have been made, open questions remain regarding how large-scale ensemble synchrony, cultural variations, and long-term training shape the neural computations that underpin coordinated movement. Together, current evidence suggests that dance synchrony serves as a rich subject for understanding how brains and bodies coordinate to produce shared aesthetic and social experience.

To study motor and neural synchronization, the present research offers a combination of approaches to study collaborative performance through motif-based neural and motor synchronization. The study conducted by Miranda et al. (2018) investigated the formation process of complex hand movements and found that complex movements can be decomposed into a combination of simpler reaching movements, a method known as Movement Element Decomposition (MED). This decomposition revealed a scale-free relationship between mean velocity and displacement, characterized by an optimal exponent $\alpha=2/3$. Recent studies on the energy balance of human movement suggest that this $\alpha$ exponent can vary. For instance, differences have been observed when comparing the dominant and non-dominant hands of right-handed individuals during writing (Ramos et al., 2025a). Additionally, this exponent may exhibit unique alterations across different movement directions, such as when assessing balance in the mediolateral and anteroposterior directions (Fialho et al., 2025). Furthermore, Ramos et al. (2025b) demonstrated that the value of the $\alpha$ exponent varies over time during fine motor tasks, with more significant variation occurring in the dominant hand. This emphasizes the reduction of degrees of freedom (Bernstein, 1967) in motor executions with the non-dominant hand. These findings encourage us to investigate other circumstances in which this optimization index may be altered. We hypothesize that collaborative dance, as a genuinely creative motor task, may exhibit dynamics of $\alpha$ that explore differing degrees of freedom. We expect these movement dynamics to change and become interrelated. To investigate this, we will measure how the $\alpha$ index varies over time and how it may synchronize between subjects improvising together under various types of interaction stimuli.

The coordination of human movement is fundamentally defined by Bernstein's Degrees of Freedom (DoF) Problem (Bernstein, 1967), which addresses the redundancy inherent in the sensorimotor system, in which numerous joints and muscles can be orchestrated to achieve a single task goal. The central challenge for the Central Nervous System is managing this high-dimensional space of possibilities, a process typically solved through the temporary formation of synergies, functional groupings of DoF that simplify control (Latash & Turvey, 1996). Motor learning is, therefore, a progression from initially "freezing" DoF to reduce complexity toward "freeing" and "exploiting" of redundant DoF, allowing for increased flexibility and adaptability in movement (Vereijken et al., 1992). Mastery over DoF dictates the complexity and dimensionality of motor output.

We hypothesize that in interactive, creative tasks, deliberately increasing explored DoF, which reflects greater individual movement complexity, will disrupt simpler, low-dimensional coordination solutions. This disruption may lead to reduced interpersonal movement synchronization between participants, even when they are performing together.

Synchronization in collaborative dance offers a unique lens through which to explore how humans coordinate their actions to create shared aesthetic and social experiences. By bridging neuroscience, principles of motor control, and biomechanics, our study aims to deepen our understanding of how creative motor tasks like improvisational dance shape neural and motor synchronization.

This study employs an experimental approach that combines EEG measurements and movement in collaborative dance, pioneering the investigation of neural synchronization during the process of group improvisation in dance.

## 1.1 From a physical perspective:

**(i)** Collaborative improvisation can be framed as a high-dimensional complex system in which neural oscillations and motor trajectories constitute coupled dynamical variables evolving across multiple temporal scales. Such multiscale interactions are characteristic of biological coordination and other high-dimensional dynamical systems operating far from equilibrium (Haken et al., 1983; Kelso, 1995).

**(ii)** Fluctuations in movement strategies, captured here through the α-exponent, resemble local power-law relations commonly found in scale-invariant processes (Bak, 1996; Stanley et al., 1999). Likewise, motif-based neural synchronization characterizes transient coherence patterns analogous to metastable states in coupled oscillator networks (Breakspear et al., 2010; Rabinovich et al., 2008). These metastable regimes have been proposed as key computational principles in neural and cognitive dynamics.

**(iii)** The collective behavior emerging from these interactions may thus be interpreted as spontaneous organization within a nonlinear multilayer network, where coordination does not arise from fixed coupling rules but from continuous transitions among attractor-like states (Arenas et al., 2008; Boccaletti et al., 2014). This theoretical bridge positions improvisational dance as a natural experiment for studying self-organization, criticality, and information flow in complex adaptive systems (Bar-Yam, 2004; Pikovsky et al., 2001), offering a physical lens for interpreting the neuro-motor signatures of *togetherness*.

## 2. Experimental Protocol and Data Acquisition

The experiment was conducted in two sessions: before and after a movement interaction program called Generative Dance through Agent Modelling (GDAM). This program translates computational models into scores for collective improvisation, and emphasizes principles of complexity theory (Leitão et al. 2025). Participants were instructed to move freely, utilizing only two basic movement states: undulating (wavy) or rectilinear (straight-line) movements with both hands.

Each experimental session involved four participants. The participants were seated at the four sides of a table, with written movement instructions placed in front of them.

**Motion Capture:** Four participants wore gloves equipped with infrared markers on both hands. Movement was captured using an OptiTrack Motion Capture system featuring 17 cameras, providing 3D spatial data with 1mm precision and a sample rate of 120Hz. The origin of the coordinate system was positioned on the floor and aligned with the center of the table. See Fig. 1 for an illustration of the setup.

**Electroencephalography (EEG):** Simultaneously, neural activity was recorded. Due to resource constraints, only two of the four participants wore EEG caps during each experimental session. EEG recordings were obtained using a 64-channel system (Neuvo 64-channel Amplifier - Compumedics), following the International 10-20 System. The Cz electrode served as the reference, and data were acquired at a sampling rate of 1000 Hz. 64 channels were divided into two caps, excluding non-cortical regions, configured to record 28 channels per individual, covering the following locations: FP1, FP2, F3, F4, F5, F7, F8, Fz, FC1, FC2, T3, T4, T5, T6, C3, C4, Cz, CP1, CP2, CP5, CP6, P3, P4, Pz, P9, P10, O1, O2, Oz.

For data processing, EEGLAB, a MATLAB toolbox, was utilized. The data were filtered using a bandpass filter between 0.5 and 48 Hz. The continuous data were then segmented into 1.28 s epochs. Subsequently, epochs in which the signal exceeded a threshold of +/- 100 μV were automatically rejected as containing artifacts. Following these automated procedures, a final visual inspection was performed, and any remaining artifact-contaminated epochs were manually removed.

The tasks were performed in the following order:

**1. (FM) Free Movement (2 min):** Participants were instructed to move purely according to their own volition, without constraint or consideration for the other participants. This task served as an individual baseline for unconstrained motor dynamics.

**2. (DM) Dependent Movement (2 min):** In this phase, participants were encouraged to actively consider the movements of the other participants and decide how they would integrate or respond to these movements. The specific nature of the interaction (e.g., imitation, opposition, synchronization) was left entirely to the subjects' choice.

**3. (RM) Rule-Based Movement (6 min):** This task introduced specific, structured rules (Wolfram rules, see Fig. 2.), see for movement, effectively simulating a cellular automaton algorithm. Movement became dependent on the state (movement states: undulating or rectilinear) of a participant's neighbors. Initial State: Each participant was given an initial movement configuration (either rectilinear or undulating). Rule Changes: The period of the rule was 30 seconds. At the end of each period, the "**observe**" command was given, during which participants had 10 seconds to identify the neighbors' configuration on the rule sheet and understand what their next configuration would be. The observing process was carried out continuously, without interrupting the movement. After that, the "**act**" command was given, which corresponded to the transition to the next rule, which was then followed until the next "**act**" command.

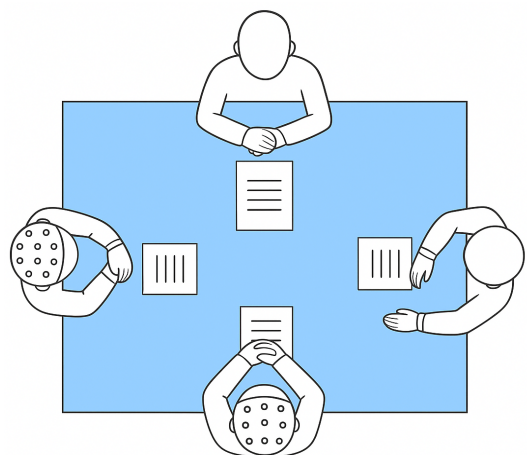

**Fig. 1.** Protocol setup illustration, including two subjects wearing EEG caps. Sheets on the table contain transition rules such as those in Fig 2.

## 2.1 Wolfram rules

In the RM task, 4 Wolfram rules were used for two-dimensional cellular automata with four cells. Formally, let n denote the number of binary cells (where 0 or 1 represents a pattern of undulating or rectilinear movements) in a local neighborhood, here n = 4: the focal cell $\sigma$ and its three neighbors $\sigma_r, \sigma_f, \sigma_l$ (right, front and left respectively). A transition rule is completely specified by the $2^n$ bits $d_k = \varphi(b_k)$, where $\{b_k; k=0\ldots 2^n-1\}$ is the lexicographic ordering of all length-$n$ binary vectors. For example:

$b_0 = \{0,0,0,0\}$,
$b_1 = \{0,0,0,1\}$,
…,
$b_{(2^n-1)} = b_{15} = \{1,1,1,1\}$

Each integer N (N represents a code of a Wolfram rule) can be written as a sequence of binary digits $d_k$, each representing the next state for a specific neighborhood configuration $b_k$. The Wolfram integer code $N_\varphi$ associated with the rule $\varphi$ is given by the binary-to-integer map:

$$N_\varphi = \sum_{k=0}^{2^n-1} d_k \cdot 2^k$$

So that the binary expansion of $N_\varphi$ recovers the truth table of $\varphi$ bit by bit. Conversely, given an integer $N$ with binary digits $d_k$ (where $d_k \in \{0,1\}$).

This one-to-one correspondence endows the finite set of Boolean local maps. Under bitwise comparison rules, it permits systematic enumeration and algebraic manipulation of rule classes (in our case, undulating or rectilinear movements, which place nonlinear constraints on the $b_k$). In the specific 2-D diamond-neighborhood model used for choreographic objects, the case *n = 4* yields distinct local rules and makes the Wolfram code a convenient scalar index for exploring properties such as symmetric attractors, reversible cycles, and sensitivity to perturbation (Leitão et al., 2023), giving an application to emergent collective dynamics.

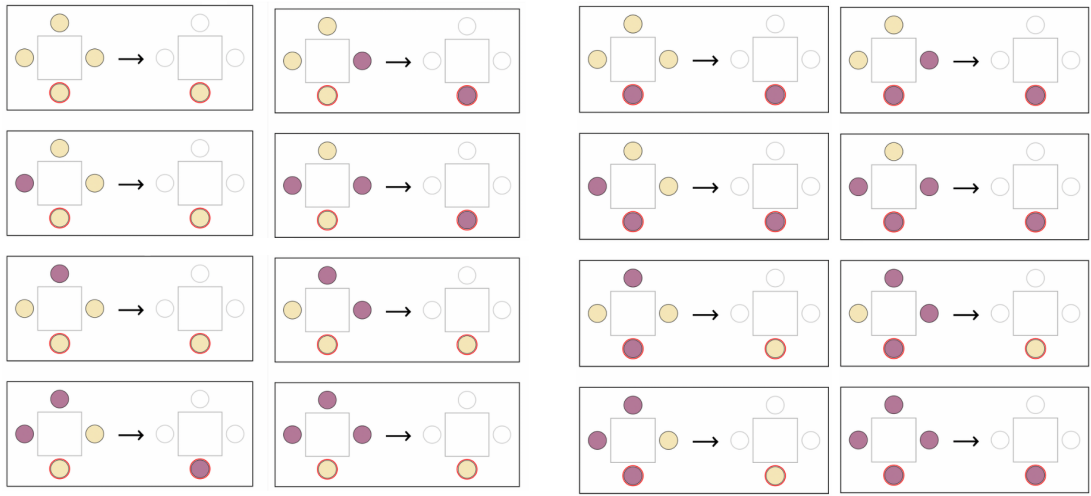

**Fig. 2.** Wolfram rule 37102 for configurations of being undulating (dusty purple) or rectilinear (cream) in the next step as a function of their neighbours (Leitão et al., 2023).

## 3. Biomechanical data segmentation and processing

Based on the work of Miranda et al. (2018), it has been shown that there exists an exponent linking mean velocity and displacement in reaching movements $\langle V \rangle \propto D^\alpha$. This relationship can be extended to more complex movements by decomposing the velocity components into zero-crossings, thereby generating sub-movements that preserve the same dynamic properties. The exponent α is an index associated with the energetic dynamics of movement and can be interpreted as an indicator of the motor strategy employed.

The methodology proposed in this study aims to identify synchronized motor patterns among different participants. To this end, the entire movement is considered for the computation of the α index, in contrast to previous approaches (Fialho et al., 2025; Miranda et al., 2018; Ramos et al., 2025a, 2025b). This work introduces a single α index derived from the combined movement of two distinct body parts, in our case, the left and right hands, as illustrated in Fig. 3. This approach aims to represent movement as a whole, encompassing the three components of three-dimensional movement.

The computation of the exponent α was based on a time series approach, as developed by Ramos et al. (2025b), who introduced a temporal sequence of α values to infer temporal and sequential variations of this parameter during task execution. In the present study, α is computed within 20-second windows using a sliding window method with frame-by-frame updates, according to the following formula:

$$\alpha(t) = (\alpha_{1\_20f},\ \alpha_{2\_(20f+1)},\ \ldots,\ \alpha_{(n-20f)\_n}) \qquad (1)$$

Where n represents the total number of recorded movement frames, and f is the sampling rate. $\alpha_{i\_j}$ represents the value of α is computed from the movement occurring between frames i and j. From the resulting α time series, we extract features that enable the analysis of motor behavior synchrony between participants on a pairwise basis.

The $\alpha$ exponent is related to the energy balance of the system, being a highly important index for studying motor dynamics, unlike velocity or acceleration, which do not represent the energy dynamics in their entirety, as they do not take into account the temporal component of the movement (Ramos et al., 2025b).

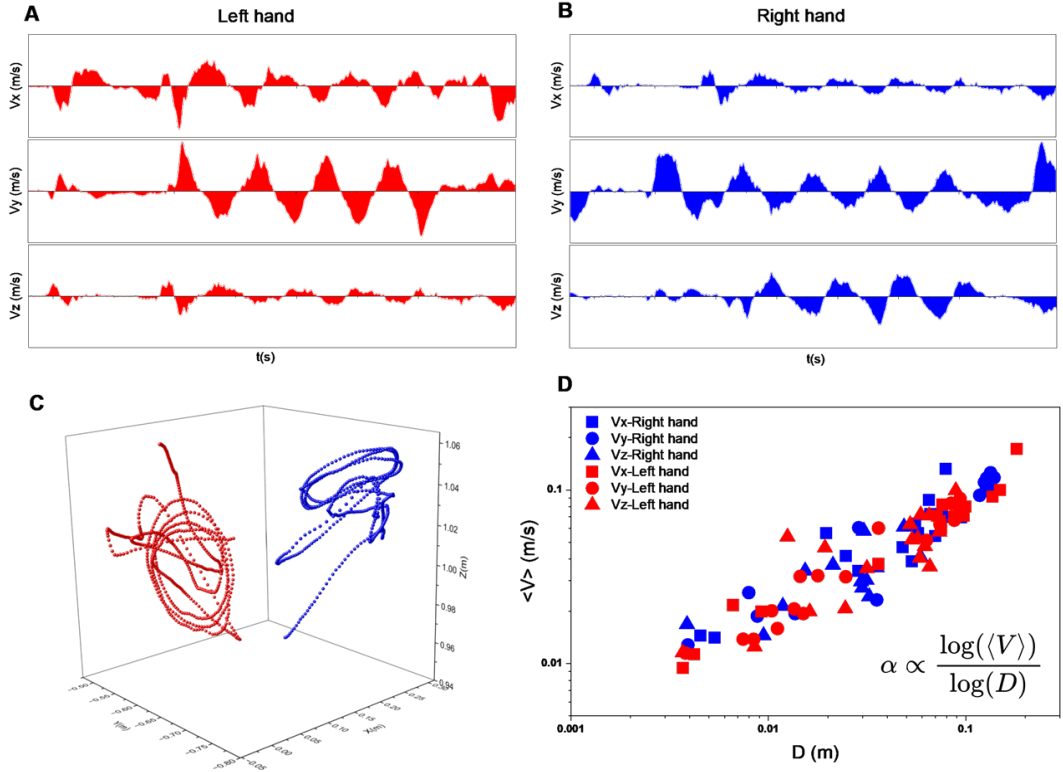

**Fig. 3.** Application of the MED method to the motion analysis of both hands combined. (**A**) and (**B**) show the velocity components over time for the left (red) and right (blue) hand, respectively. (**C**) illustrates the 3D spatial trajectories from which these velocity data were derived. (**D**) demonstrates the power-law relationship between <V> and D for each sub-movement, presented on a log-log scale.

**3. Motif Synchronization**

The study of synchronization has gained substantial traction in recent decades, particularly through the use of the Kuramoto model (Kuramoto, 1975, 1984), which has become a canonical framework for describing collective behavior in populations of coupled phase oscillators. Numerous extensions of the model have been proposed to account for heterogeneity, network structure, and time delays (Acebrón et al., 2005; Arenas et al., 2008; Rodrigues et al., 2016). These formulations have enabled profound theoretical insights into phase locking, critical transitions, and emergent coherence in complex systems. Despite its analytical power, however, the Kuramoto model presents intrinsic limitations when applied directly to real-world time series generated by interacting individuals.

In empirical settings, especially those involving biological or behavioral data from multiple persons, synchronized activity may be driven by shared sensory input, contextual cues, or global environmental fluctuations. Since the Kuramoto model assumes a predefined coupling function and does not infer influence or directionality from the data, instantaneous or near-zero-lag synchrony may reflect external common drivers rather than genuine inter-individual interaction (Breakspear et al., 2010; Pikovsky et al. 2001). Even in delay-coupled variants of Kuramoto dynamics (Earl & Strogatz, 2003; Yeung & Strogatz, 1999), time delays are modeled, not estimated, and the framework remains fundamentally phenomenological rather than data-driven. Additionally, extracting phase information from noisy, broadband, or non-stationary signals, such as EEG or behavioral time series, introduces methodological ambiguities that further compromise the interpretability of synchronization measures based on the Kuramoto model (Kralemann et al., 2008).

For these reasons, and given our primary objective of characterizing how individuals influence one another within collective dynamics, we adopt an alternative approach: the Motif-Synchronization method introduced by Rosário et al. (2015), which has been applied in a wide range of contexts for synchronized time series analysis, from the progression of epidemics to neural dynamics (Fernandes et al., 2024; de O. Toutain et al., 2023; Saba et al., 2022; Sousa et al., 2024; Thaise et al., 2024; Toutain et al., 2020; ). Instead of relying on phase representations, Motif-Synchronization transforms each time series into an ordinal sequence of three-point motifs that encode local patterns of increases and decreases in the signal. Synchronization is then quantified by computing the proportion of identical motifs occurring across two time series at specific positive time lags, which allows for the detection of lead-lag relationships and directional information flow (see Fig. 4). By evaluating this similarity over time, we construct Time-Varying Graphs (TVGs).

A connection between two nodes is established whenever the proportion of identical motifs exceeds a similarity threshold ( $S_c$ ). In the present study, we set $S_c = 0.7$ as a standard criterion, meaning that at least 70% of the observed motif transitions must be structurally identical for two time series to be considered synchronized. This threshold reflects a stringent regime of structural coincidence, well above chance levels, ensuring that detected links represent statistically meaningful synchronization rather than random overlap. Robustness analyses confirmed that the global network topology and all reported results remain qualitatively stable for thresholds in the range ( $S_c \in [0.5, 0.9]$ ), indicating that the choice of $S_c = 0.7$ does not bias the conclusions.

This framework is particularly advantageous in empirical contexts where zero-lag synchrony may arise from shared perceptual events or global inputs rather than genuine inter-individual coupling. By explicitly incorporating temporal delays in the motif-matching process, the method distinguishes between trivial stimulus-driven synchrony and temporally structured dependencies that more accurately reflect inter-individual influence. As a result, it provides a more reliable account of information flow between participants while avoiding the confounds inherent in phase-based or model-assumed synchronization frameworks.

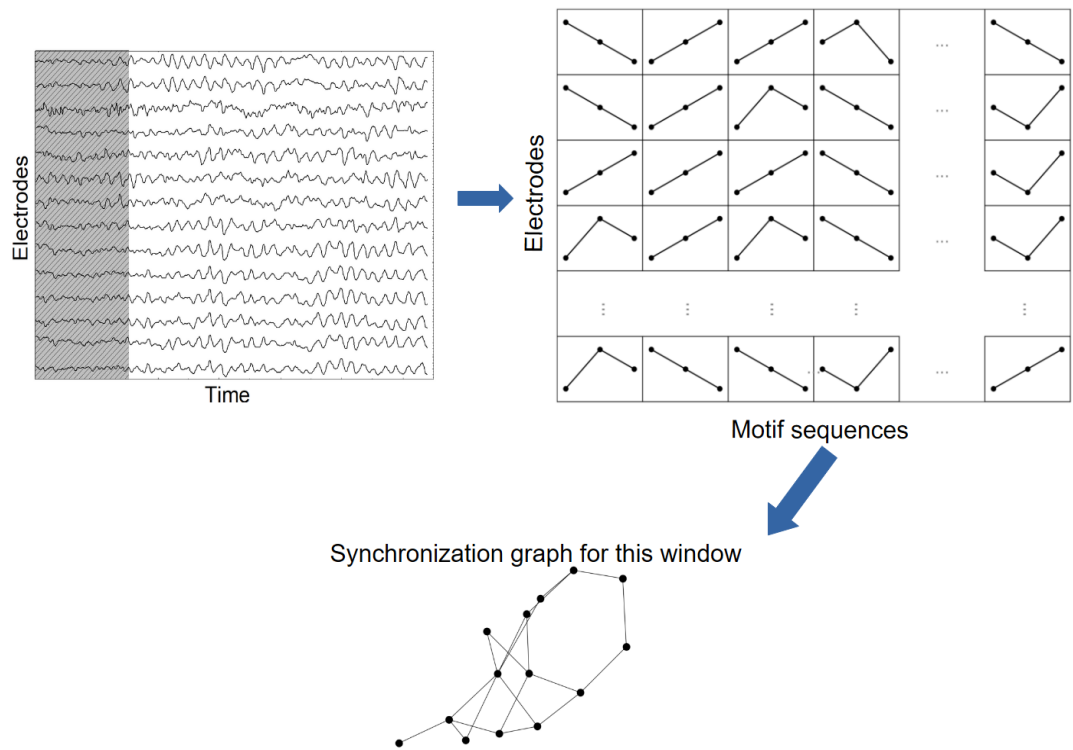

**Fig. 4.** Motif synchronization method in neural signals for a certain window. The process must be repeated window by window in a sliding sequence to obtain a TVG.

### 3.1 Biomechanical synchronization

Biomechanical synchronization was measured from the subjects' $\alpha$ time series, using motif synchronization. We sought the sequence of correlated motifs in 1s windows (120 frames), considering delays of up to 4 frames, thereby obtaining a synchronization graph for each time frame. In the end, we produced a composite graph containing all pairwise synchronizations between individuals, as shown in Fig. 5. Because our time scales are significantly shorter than those for information processing and motor response generation, we treat the graphs as undirected.

From this 6-edge graph, we analyzed the proportion of synchronized moments of that same edge (pair of individuals) throughout all tasks.

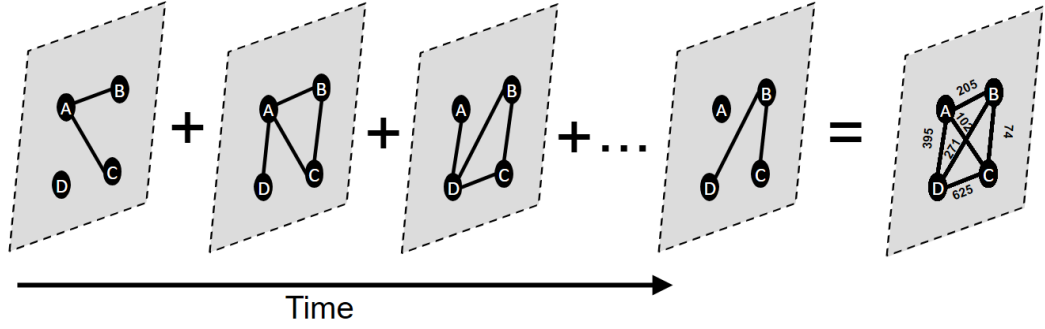

**Fig. 5.** Biomechanical synchronization graph in a α time series. Each node represents a player in Fig. 1.

### 3.2 Multilayer synchronization in neural data

To assess neural synchronization between pairs of individuals, a combined methodological approach was employed: Motif Synchronization and the multilayer TVG analysis proposed by Sousa et al. (2024). This work presents a methodology for constructing TVGs based on the quasi-simultaneous counting of micropatterns, referred to as motifs, among all possible pairs of EEG signals from each subject. This approach is considered appropriate for representing neural activity, given that the brain exhibits complex, continuous, and connectionist dynamics even under minimal external stimulation, such as during the resting state.

The study by Sousa et al. (2024) presents an innovative approach that characterizes synchronization between pairs of TVGs by identifying simultaneous connections in the networks constructed for different individuals. The method relies on the Incidence-Fidelity ($IF_{ij}$) index, which is computed as the product of the Incidence index, representing the number of occurrences of the same edge in both TVGs divided by the total time instants, and the Fidelity index, which represents the proportion of times an edge appears simultaneously in both TVGs relative to the total number of its occurrences. Mathematically, the Incidence-Fidelity index for two TVGs, G and F, is expressed as follows:

$$IF_{i,j} = I_{i,j} \times F_{i,j} \qquad (2)$$

,where:

$$I_{i,j} = \frac{\sum_{s=1}^{\Gamma} a_{i,j}^{G}(t_s) a_{i,j}^{F}(t_s)}{\Gamma} \qquad (3)$$

and:

$$F_{i,j} = \frac{\sum_{s=1}^{\Gamma} a_{i,j}^{G}(t_s) a_{i,j}^{F}(t_s)}{\sum_{s=1}^{\Gamma} a_{i,j}^{G}(t_s) + \sum_{s=1}^{\Gamma} a_{i,j}^{F}(t_s) - \sum_{s=1}^{\Gamma} a_{i,j}^{G}(t_s) a_{i,j}^{F}(t_s)} \qquad (4)$$

,where:

$$a_{i,j}^{x}(t_s)=\{1, if\ \exists\ an\ edge\ e_{i,j}\in theframe\ t_s\ do\ TVG\ x.\ s=1,2,\ldots\ \Gamma\ ;\ 0, otherwise\} \quad (5)$$

This method results in a weighted network, in which the weights associated with the edges correspond to the IF of each link. See Fig. 6.

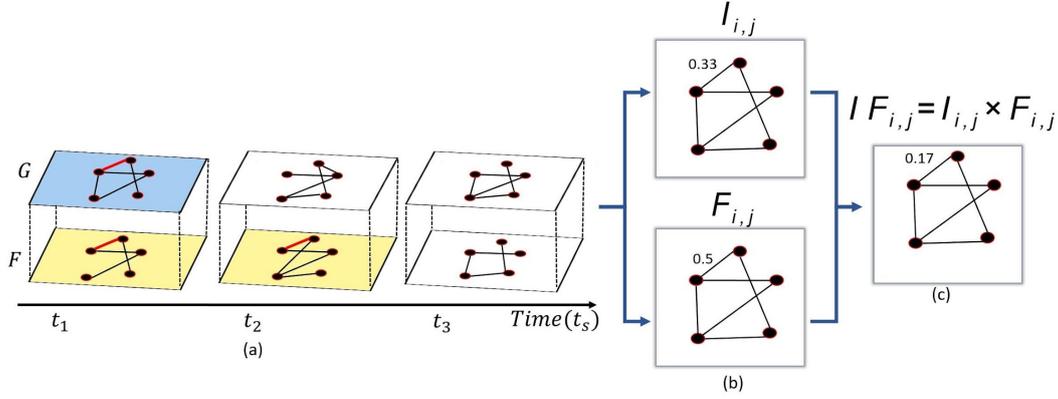

**Fig. 6.** Example of calculating synchronization indices in a multilayer network for an edge: (**a**) Multilayer network with 5 vertices in each layer and TVG lifetime equal to 3. (**b**) Fidelity and Incidence Network for the considered TVGs. (**c**) Incidence-Fidelity Network.

The Incidence-Fidelity network for a pair of TVGs aims to avoid misinterpretations of synchronization values when analyzed in isolation. For instance, an edge that occurs infrequently but is always simultaneous in both networks may show high Fidelity values, even if it appears only once throughout the TVGs. Thus, the Incidence-Fidelity index measures synchronization between networks by considering both the temporal extent of the TVGs and their individual edge frequencies, ensuring that the maximum interlayer synchronization value effectively corresponds to the highest synchronization recorded between the TVGs.

The Incidence-Fidelity networks were constructed for each pair of individuals. To assess whether the identified edges could arise by randomness, additional Incidence-Fidelity networks were generated with randomized edges for each pair of TVGs. The highest Incidence-Fidelity value obtained was adopted as only edges with weights exceeding this threshold were retained for analysis.

To compare engagement across tasks, we also examined the mean edge weight distribution of the Incidence networks and compared these distributions before and after GDAM using the Mann-Whitney statistical test.

## 4. Brain synchronization results

For the EEG data, the direct comparison between tasks before and after the GDAM showed no significant effect, $F(1, 8) = 0.216$, $p = .654$, $\eta^2_p = .026$. In contrast, the relative EEG analysis revealed a significant effect, $F(1, 8) = 10.787$, $p = .011$, $\eta^2_p = .574$, indicating a marked change in the relative neural activity patterns between tasks after the GDAM in the comparison between the tasks FM and DM (p=0.01). Post-Hoc analysis is available in Table 1.

In order to compare the mean edge incidence distributions, Mann-Whitney U tests were conducted to compare PRE (1) and POS (2) conditions across the three tasks (FM, DM, and RM). Results indicated a significant difference for FM (U = 344,717.5; Z = 6.97; p < .001) and RM (U = 402,018.0; Z = 13.74; p < .001), with the POS condition showing higher mean ranks (FM: $M_1$ = 678.52, $M_2$ = 834.48; RM: $M_1$ = 602.73, $M_2$ = 910.27). In contrast, no significant difference was observed for DM (U = 270,162.0; Z = –1.84; p = .195), although PRE exhibited slightly higher mean ranks ($M_1$ = 777.14; $M_2$ = 735.86). A Bonferroni correction was applied to control for Type I error. (See Fig. 3.).

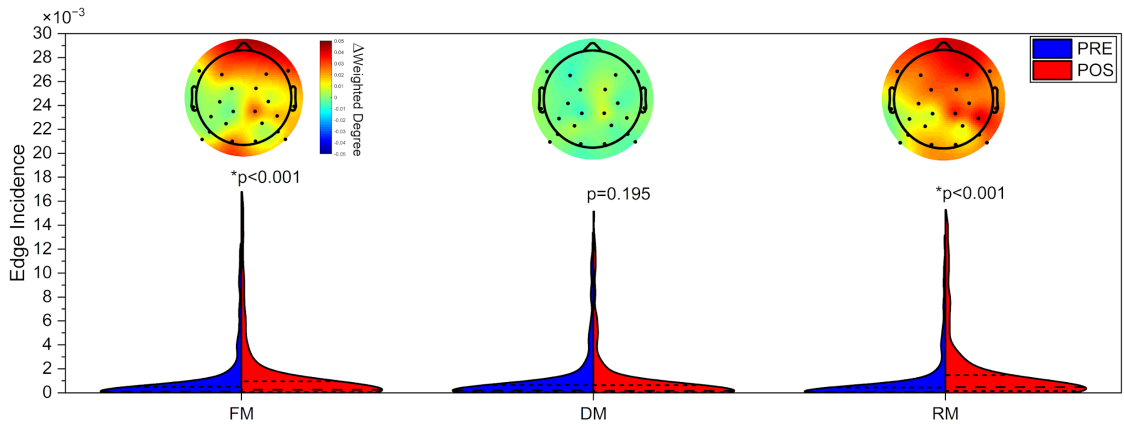

**Fig. 7.** Distribution of edge Incidence values across tasks (FM, DM, RM) before (PRE, blue) and after (POS, red) the GDAM. Violin plots represent the distribution of edge incidence values within each condition, with horizontal dashed lines indicating the median. Insets show the ΔWeighted Degree (POS-PRE) topographic maps.

The comparison of the mean edge incidence distributions revealed greater synchronization in the POS condition for the FM and RM tasks, with a prominent effect in the frontal region, as illustrated by the insets showing the variation of the weighted degree across electrodes in Fig. 7.

## 5. Biomechanical synchronization results

The statistical analyses revealed a significant effect for the biomechanical measures when comparing performance directly before and after the GDAM, $F(1, 70)$ = 5.233, $p$ = .025, $\eta^2_p$ = .070, indicating a change in biomechanical performance following the GDAM. However, when analyzing the relative changes between tasks, no significant difference was observed, $F(1, 70)$ = 0.006, $p$ = .940, $\eta^2_p$ = .000, suggesting that the proportional relationship between tasks remained consistent across sessions.

|  | Mean Difference (PRE-POS) | Std. Error | P value | 95% Confidence Interval for Difference | |
|---|---|---|---|---|---|
|  |  |  |  | Lower Bound | Upper Bound |
| **EEG (PRE: 5 pairs, POS: 5 pairs)** | | | | | |
| FM | 0.08 | 0.08 | 0.36 | -0.11 | 0.27 |
| DM | -0.06 | 0.06 | 0.37 | -0.19 | 0.08 |
| RM | -0.1 | 0.07 | 0.2 | -0.27 | 0.06 |
| **Relative increase in EEG** | | | | | |
| DM,FM | -.083* | 0.03 | 0.01 | -0.14 | -0.03 |
| RM,FM | -0.11 | 0.06 | 0.13 | -0.25 | 0.04 |
| RM,DM | -0.02 | 0.05 | 0.66 | -0.14 | 0.09 |
| **Biomechanics** (PRE: 42 pairs, POS: 30 pairs) | | | | | |
| FM | 3.099* | 1.46 | 0.04 | 0.19 | 6.01 |
| DM | 0.19 | 1.31 | 0.89 | -2.42 | 2.8 |
| RM | 3.550* | 1.46 | 0.02 | 0.64 | 6.46 |
| **Relative increase in biomechanics** | | | | | |
| DM,FM | -0.12 | 0.11 | 0.27 | -0.33 | 0.1 |
| RM,FM | 0.05 | 0.11 | 0.64 | -0.17 | 0.28 |
| RM,DM | 0.08 | 0.11 | 0.45 | -0.13 | 0.3 |

*. The mean difference is significant at the .05 level. Correction for multiple comparisons: Bonferroni.

**Table 1.** Statistical comparisons between measurements before and after the GDAM.

The post-hoc analyses revealed a decrease in motor synchronization for the FM and RM tasks after the GDAM (see Table 1 and Fig. 8). However, no significant differences were found when comparing the variation of these measures (Δ/sum) between the PRE and POS conditions, as also observed in the EEG analyses (see Fig. 9). This suggests consistency in synchronization patterns across tasks within the same day, with no significant differences between tasks when comparing PRE and POS sessions.

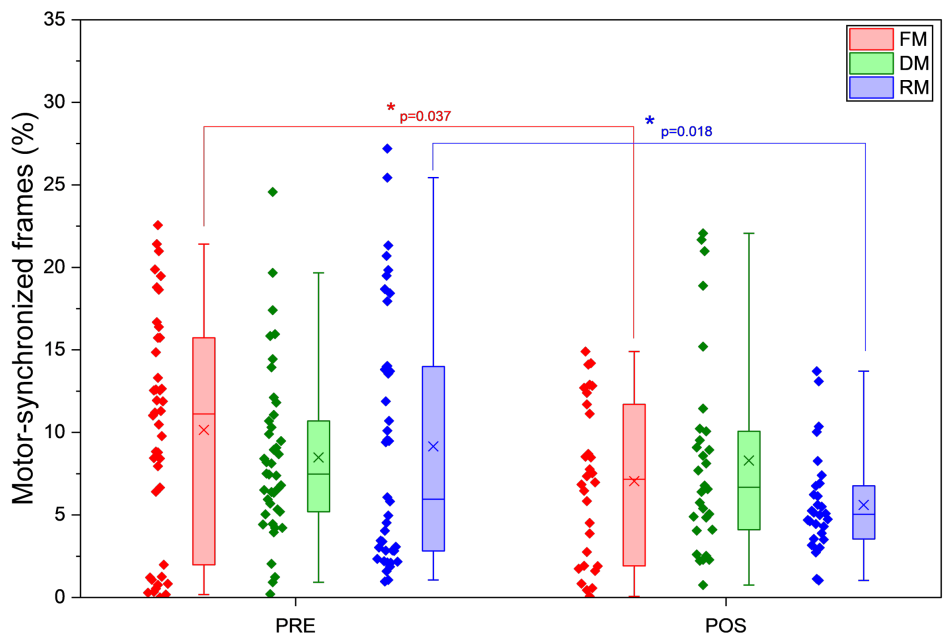

**Fig. 8.** Ratio of motor-synchronized frames between pairs before and after the GDAM.

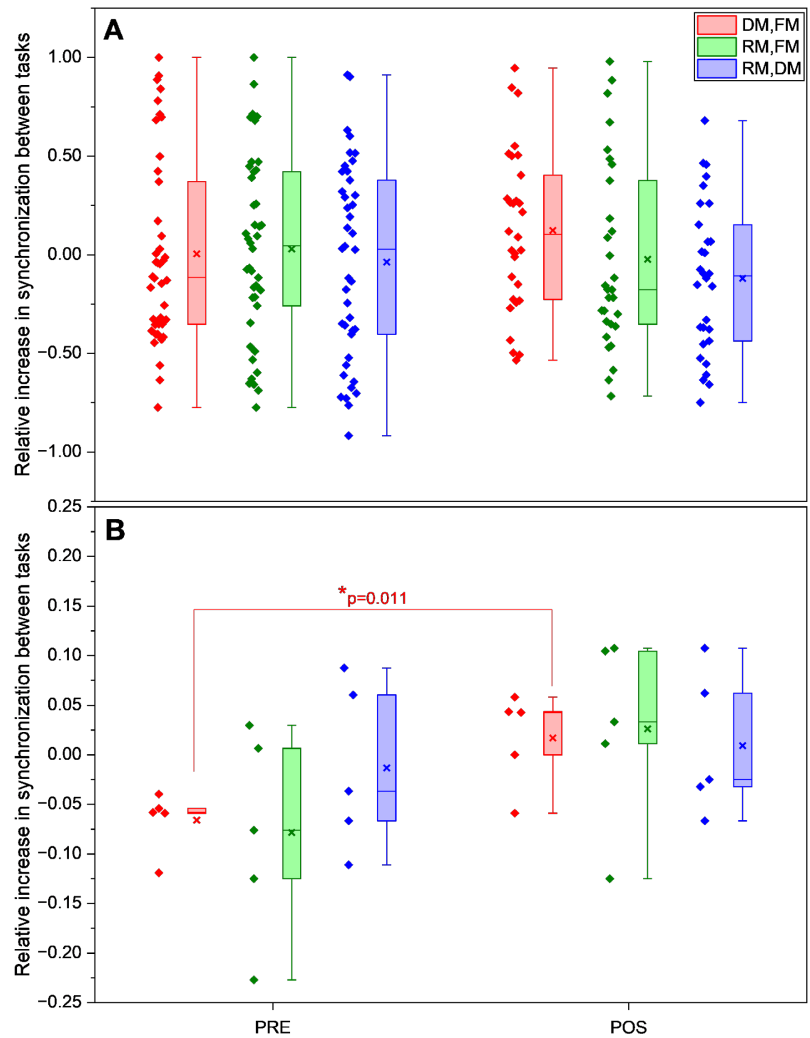

**Fig. 9.** Relative increase in synchronization between tasks, comparing pairs before and after the GDAM. (**A**): motor synchronization; (**B**) Brain synchronization.

## 6. Discussion

**Neural behavior**

The results suggest that frontal lobe synchronization between individuals does not directly reflect explicit social interaction, but rather alignment in internal executive and intentional processes. Both free improvisation and rule-based improvisation demand greater autonomous generation of motor strategies, monitoring, inhibition, and decision-making under uncertainty, elements inherent to a genuine sense of *togetherness*. These processes rely on the prefrontal cortex (Badre, 2008; Bari & Robbins, 2013; Krawczyk, 2002), especially the right prefrontal region (Aron et al., 2004, 2014). This neural engagement leads to increased inter-individual coupling in these areas. In contrast, explicit interaction conditions reduce decisional autonomy and shift cognitive focus toward sensorimotor and parietal networks, resulting in decreased shared frontal involvement and, consequently, no increase in synchronization in this region.

**Opposite trends and DoF**

The results indicate increased neural synchrony between individuals after the GDAM program, while motor synchrony decreased. A similar dissociation, this opposite effect between neural and motor behavior, was also reported by Ramos et al. (2025b), who observed higher permutation entropy in neural data during writing with the non-dominant hand, whereas permutation entropy of motor data was higher in the dominant hand. The connection between this finding and ours suggests an inverse relationship between brain and motor dynamics, which may indicate greater activity efficiency: in our case, the more synchronized the brain, the less synchronized the hands. This pattern reflects an expansion of motor degrees of freedom (a behavior of increased DoF), as participants enhanced their experience in collective improvisation, enabling a broadening of their range of possibilities during generative dance. Such expansion enables more diverse and flexible compositions, which are not necessarily identical across individuals (Himberg et. al, 2018; Tseng et. al, 2021). Despite this increase in motor variability, neural synchronization simultaneously intensified, indicating that the sense of *togetherness* is maintained even as motor behaviors diversify.

## 7. Conclusion

This study highlights modern techniques in movement analysis and synchronization modeling and is pioneering in characterizing the nonlinear dynamics underlying generative dance through simultaneously recorded neural and motor data. Our findings demonstrate that the participatory sense emerging during collaborative improvisation is supported by a complex, multilayer network of neural interactions shaped by training. These interactions reveal signatures of metastability, self-organization, and coupling, indicating that *togetherness* emerges not from simple motor imitation, but from coordinated transitions across attractor-like states in shared neuro-cognitive space.

Importantly, the observed dissociation between increased inter-brain synchronization and reduced motor synchrony after training suggests a reorganization of the system's degrees of freedom: performers expand the dimensionality of their motor manifolds even as their neural dynamics become more coherent. This coupling-decoupling paradox is a hallmark of complex adaptive systems, in which global coherence coexists with local divergence. This dynamic can be situated within Araújo's (2009) ecological model of decisional behavior, particularly its third phase of action calibration within the perceptual–motor system. The post-training dissociation between neural and motor synchrony indicates that performers operate in this calibration phase, fine-tuning their actions in relation to shared intentions and contextual affordances rather than converging on uniform motor outputs. Increased neural coherence thus reflects stabilized collective intentionality, while motor divergence expresses the adaptive expansion of individual action possibilities. Thus, *togetherness* in generative dance cannot be reduced to aligned or mirrored movements; rather, it reflects an emergent property of a multimodal, dynamical system in which neural intentionality, creative exploration, and interpersonal influence continuously reorganize.

By integrating physical concepts such as metastability, scale-invariant structure, and nonlinear network dynamics into the study of collective improvisation, this work positions dance as an experimental model for examining self-organization and information flow in adaptive systems. Beyond the artistic domain, these findings provide a conceptual and methodological framework for investigating coordination and emergent collective behavior in broader biological, cognitive, and social systems.

**Funding**

National Council for Scientific and Technological Development (CNPq) and Brazilian Federal Agency for Support and Evaluation of Graduate Education (CAPES) provided financial support for this project: Jose Garcia Vivas Miranda (grant CNPq nº 308758/2021-8), Adriana de Faria Gehres (grant CNPq nº 420222/2022-7) and Yago Emanoel Ramos (PhD scholarship CAPES nº 88887.203818/2025-00). Additional funding was provided by the Portuguese Foundation for Science and Technology (FCT): Maria João Alves (FCT Multiannual Funding UIDB/00472/2020) and Ana Maria Leitão (FCT PhD grant 2021.07216.BD).

**Ethics approval**

The experimental protocol was approved by the Ethics Committee of the Physics Institute of Federal University of Bahia, Certificate of Presentation for Ethical Approval: 73085523.0.0000.8035, in accordance with the Declaration of Helsinki.

**CRediT**

**Yago Emanoel Ramos:** Writing - original draft, Investigation, Data Curation, Formal Analysis, Software, Conceptualization, Methodology; **Raphael Rosário:** Writing - original draft, Software, Conceptualization, Formal Analysis, Validation, Investigation, Methodology; **Adriana de Faria Gehres:** Writing - review & editing, Supervision, Methodology, Conceptualization, Project administration; **Maria João Alves:** Writing - review & editing, Conceptualization, Formal Analysis, Methodology, Validation, Investigation, Supervision; **Ana Maria Leitão:** Conceptualization, Validation, Investigation, Methodology; **Cecília Bastos da Costa Accioly:** Writing - review & editing, Methodology, Conceptualization; **Fatima Wachowicz:** Writing - review & editing, Supervision, Methodology, Conceptualization, Investigation; **Ivani Lúcia Oliveira de Santana:** Writing - review & editing, Methodology; **José Garcia Vivas Miranda:** Writing - review & editing, Data Curation, Supervision, Methodology, Software, Formal Analysis, Validation, Investigation, Conceptualization, Resources.

**Declaration of competing interest**

The authors declare no conflict of interest.

## Data availability

Data will be made available on request.

## References

Acebrón, J. A., Bonilla, L. L., Vicente, C. J. P., Ritort, F., & Spigler, R. (2005). The Kuramoto model: A simple paradigm for synchronization phenomena. *Reviews of Modern Physics, 77*(1), 137–185. https://doi.org/10.1103/RevModPhys.77.137

Araújo, D. (2009). O desenvolvimento da competência táctica no desporto: O papel dos constrangimentos no comportamento decisional. *Motriz: Revista de Educação Física*, 15(3), 537–540. https://doi.org/10.5016/2942

Arenas, A., Díaz-Guilera, A., Kurths, J., Moreno, Y., & Zhou, C. (2008). Synchronization in complex networks. *Physics Reports, 469*(3), 93–153. https://doi.org/10.1016/j.physrep.2008.09.002

Aron, A. R., Robbins, T. W., & Poldrack, R. A. (2004). Inhibition and the right inferior frontal cortex. *Trends in cognitive sciences, 8*(4), 170-177.

Aron, A. R., Robbins, T. W., & Poldrack, R. A. (2014). Right inferior frontal cortex: addressing the rebuttals. *Frontiers in Human Neuroscience, 8*, 905. https://doi.org/10.3389/fnhum.2014.00905

Badre, D. (2008). Cognitive control, hierarchy, and the rostro–caudal organization of the frontal lobes. *Trends in cognitive sciences, 12*(5), 193-200.

Bak, P. (2013). *How nature works: The science of self-organized criticality*. Springer Science & Business Media.

Bar-Yam, Y. (2014, February). Complex Systems Science: From Cell Regulation to the Global Food Crisis. In *ISCS 2013: Interdisciplinary Symposium on Complex Systems* (pp. 19-28). Berlin, Heidelberg: Springer Berlin Heidelberg.

Bari, A., & Robbins, T. W. (2013). Inhibition and impulsivity: Behavioral and neural basis of response control. *Progress in neurobiology*, *108*, 44-79.

Basso, J. C., Satyal, M. K., & Rugh, R. (2021). Dance on the brain: enhancing intra-and inter-brain synchrony. *Frontiers in human neuroscience, 14*, 584312.

Bernstein, N. (1967). *The co-ordination and regulation of movements*. Oxford Pergamon.Boccaletti, S., Bianconi, G., Criado, R., Del Genio, C. I., Gómez-Gardenes, J., Romance, M., Sendiña-Nadal, I., Wang, Z., & Zanin, M. (2014). The structure and dynamics of multilayer networks. *Physics Reports, 544*(1), 1-122. https://doi.org/10.1016/j.physrep.2014.07.001

Breakspear, M., Heitmann, S., & Daffertshofer, A. (2010). Generative models of cortical oscillations: From Kuramoto to the nonlinear Fokker–Planck equation. *Frontiers in Human Neuroscience, 4*, 190. https://doi.org/10.3389/fnhum.2010.00190

Chauvigné, L. A. S., & Brown, S. (2018). Role-specific brain activations in leaders and followers during joint action. *Frontiers in Human Neuroscience, 12,* 401. https://doi.org/10.3389/fnhum.2018.00401

Chauvigné, L. A. S., Belyk, M., & Brown, S. (2018). Taking two to tango: fMRI analysis of improvised joint action with physical contact. *PLoS ONE, 13,* e0191098. https://doi.org/10.1371/journal.pone.0191098

Cross, E. S., Darda, K. M., Moffat, R., Muñoz, L., Humphries, S., & Kirsch, L. P. (2024). Mutual gaze and movement synchrony boost observers' enjoyment and perception of togetherness when watching dance duets. *Scientific Reports, 14*(1), 24004. https://doi.org/10.1038/s41598-024-72659-7

De Jaegher, H., & Di Paolo, E. (2007). Participatory sense-making: An enactive approach to social cognition. *Phenomenology and the cognitive sciences, 6*(4), 485-507.

de O. Toutain, T. G. L., Miranda, J. G. V., do Rosário, R. S., & de Sena, E. P. (2023). Brain instability in dynamic functional connectivity in schizophrenia. *Journal of Neural Transmission, 130*(2), 171-180. https://doi.org/10.1007/s00702-022-02579-1

Earl, M. G., & Strogatz, S. H. (2003). Synchronization of identical oscillators with time delay. *Physical Review E, 67*(3), 036204. https://doi.org/10.1103/PhysRevE.67.036204

Fernandes, L. A., Apolinário-Souza, T., Castellano, G., Fortuna, B. C., & Lage, G. M. (2024). Hand differences in aiming task: A complementary spatial approach and analysis of dynamic brain networks with EEG. *Behavioural Brain Research, 469,* 114973. https://doi.org/10.1016/j.bbr.2024.114973

Gesbert, V., Hauw, D., Kempf, A., Blauth, A., & Schiavio, A. (2022). Creative togetherness. A joint-methods analysis of collaborative artistic performance. *Frontiers in psychology, 13*, 835340. https://doi.org/10.3389/fpsyg.2022.835340

Haken, H., Cardona, M., Fulde, P., & Queisser, H.-J. (Eds.). (1983). *Springer series in synergetics* (Vol. 269). Springer.

Himberg, T., Laroche, J., Bigé, R., Buchkowski, M., & Bachrach, A. (2018). Coordinated interpersonal behaviour in collective dance improvisation: the aesthetics of kinaesthetic togetherness. *Behavioral Sciences, 8*(2), 23. https://doi.org/10.3390/bs8020023

Hove, M. J., & Risen, J. L. (2009). It's all in the timing: Interpersonal synchrony increases affiliation. *Social Cognition, 27*(6), 949–961. https://doi.org/10.1521/soco.2009.27.6.949

Kelso, J. S. (1995). *Dynamic patterns: The self-organization of brain and behavior*. MIT Press.

Kralemann, B., Cimponeriu, L., Rosenblum, M., Pikovsky, A., & Mrowka, R. (2008). Phase dynamics of coupled oscillators reconstructed from data. *Physical Review E, 77*(6), 066205. https://doi.org/10.1103/PhysRevE.77.066205

Kuramoto, Y. (1975). Self-entrainment of a population of coupled nonlinear oscillators. In H. Araki (Ed.), *International Symposium on Mathematical Problems in Theoretical Physics* (pp. 420–422). Springer.

Kuramoto, Y. (1984). *Chemical oscillations, waves, and turbulence*. Springer.

Krawczyk, D. C. (2002). Contributions of the prefrontal cortex to the neural basis of human decision making. *Neuroscience & Biobehavioral Reviews, 26*(6), 631-664. https://doi.org/10.1016/s0149-7634(02)00021-0

Latash, M. L., & Turvey, M. T. (Eds.). (1996). *Dexterity and its development*. Lawrence Erlbaum Associates.

Leitão, A. M., & Alves, M. J. (2024). An essay on the dimensions of generative dance. *International Journal of Performance Arts and Digital Media, 20*(3), 421–445. https://doi.org/10.1080/14794713.2024.2357826

Leitão, A. M., Severino, R. J. M., & Alves, M. J. (2025). Generative dance: Interplaying to compose collectively. *Journal of Dance Education.* Advance online publication. https://doi.org/10.1080/15290824.2025.2565760

Fialho, K. L., Miranda, J. G. V., Ramos, Y. E., & Lucena, R. C. S. (2025). Characterization of neurophysiological, motor, and emotional biomarkers in adolescents with ASD: An integrated analysis with qEEG, facial expression, and biomechanics analysis. *Clinical EEG and Neuroscience*. Advance online publication. https://doi.org/10.1177/15500594251394773

Miranda, J. G. V., Daneault, J.-F., Vergara-Diaz, G., Torres, Â. F. S. de O. e, Quixadá, A. P., Fonseca, M. de L., Vieira, J. P. B. C., Santos, V. S. dos, Figueiredo, T. C. da, Pinto, E. B., Peña, N., & Bonato, P. (2018). Complex upper-limb movements are generated by combining motor primitives that scale with the movement size. *Scientific Reports, 8,* 12918. https://doi.org/10.1038/s41598-018-31166-6

Noy, L., Levit-Binun, N., & Golland, Y. (2015). Being in the zone: physiological markers of togetherness in joint improvisation. *Frontiers in human neuroscience, 9*, 187.

Orgs, G., Vicary, S., Sperling, M., Richardson, D. C., & Williams, A. L. (2024). Movement synchrony among dance performers predicts brain synchrony among dance spectators. *Scientific Reports, 14*(1), 22079. https://doi.org/10.1038/s41598-024-73438-0

Pikovsky, A., Rosenblum, M., & Kurths, J. (2001). *Synchronization: A universal concept in nonlinear sciences*. Cambridge University Press.

Rabinovich, M. I., Varona, P., Selverston, A. I., & Abarbanel, H. D. (2006). Dynamical principles in neuroscience. *Reviews of modern physics, 78*(4), 1213-1265. https://doi.org/10.1103/RevModPhys.78.1213

Ramos, Y. E., Santos, M. T., Yamamoto, I. N. R., da Costa Accioly, C. B., Daneault, J. F., de Almeida Filho, D. G., & Miranda, J. G. V. (2025). Handedness and brain lateralization: A nonlinear motor approach combined with EEG. *Human Movement Science, 104*, 103425. https://doi.org/10.1016/j.humov.2025.103425

Ramos, Y. E., Torres, Â. F., da Costa Accioly, C. B., Matias, F. S., & Miranda, J. G. V. (2025). Linking biomechanical model dynamics and neural complexity: Permutation entropy approaches to motor control. *Chaos, Solitons & Fractals, 201*, 117412. https://doi.org/10.1016/j.chaos.2025.117412

Ribeiro, H. L., Caixeta, F. V., Venâncio, P. E. M., Monteiro, L. C. P., & Ramos, I. A. (2024). Neuroestética da dança: uma análise neurobiológica do dançarino ao espectador. *Revista Pensar a Prática, 27*, e.76031. https://doi.org/10.5216/rpp.v27.76031

Rodrigues, F. A., Peron, T. K. D., Ji, P., & Kurths, J. (2016). The Kuramoto model in complex networks. *Physics Reports, 610*, 1–98. https://doi.org/10.1016/j.physrep.2015.10.008

Rosário, R. S., Cardoso, P. T., Muñoz, M. A., Montoya, P., & Miranda, J. G. V. (2015). Motif-Synchronization: A new method for analysis of dynamic brain networks with EEG. *Physica A, 439*, 7–19. https://doi.org/10.1016/j.physa.2015.07.018

Saba, H., Nascimento Filho, A. S., Miranda, J. G., Rosário, R. S., Murari, T. B., Jorge, E. M., ... & Araújo, M. L. (2022). Synchronized spread of COVID-19 in the cities of Bahia, Brazil. Epidemics, 39, 100587.

Severino, R., Leitão, A. M., & Alves, M. J. (2024, July). On the Dynamics of a Family of 2D Finite Cellular Automata. In *International Conference on Computational Science and Its Applications* (pp. 263-273). Springer Nature Switzerland.

Sousa, R. A., Vasconcelos, R. N., Rosário, R. S., & Miranda, J. G. V. (2024). Multilayer time-variant network edges synchronization method. *Physica D: Nonlinear Phenomena, 466*, 134219. https://doi.org/10.1016/j.physd.2024.134219

Stanley, H. E. (1999). Scaling, universality, and renormalization: Three pillars of modern critical phenomena. *Reviews of modern physics, 71*(2), S358.

Thaise, G. D. O., Miranda, J. G. V., do Rosário, R. S., & de Sena, E. P. (2024). Directed brain interactions over time: A resting-state EEG comparison between schizophrenia and healthy individuals. *Psychiatry Research: Neuroimaging, 344*, 111861. https://doi.org/10.1016/j.pscychresns.2024.111861

Toutain, T. G., Baptista, A. F., Japyassú, H. F., Rosário, R. S., Porto, J. A., Campbell, F. Q., & Miranda, J. G. V. (2020). Does meditation lead to a stable mind? Synchronous stability and time-varying graphs in meditators. *Journal of Complex Networks, 8*(6), cnaa049. https://doi.org/10.1093/comnet/cnaa049

Tseng, C. H., Cheng, M., Matout, H., Fujita, K., Kitamura, Y., Shioiri, S., & Bachrach, A. (2021). Perceived "togetherness" and "MA" between two dancers in joint improvisation. [Preprint]. HAL. https://hal.archives-ouvertes.fr/ffhal-03103040f

Vereijken, B., van Emmerik, R. E., Whiting, H. T., & Newell, K. M. (1992). Freezing degrees of freedom in skill acquisition. *Journal of Motor Behavior, 24*(1), 133-142. https://doi.org/10.1080/00222895.1992.9941608

Vicary, S., Sperling, M., von Zimmermann, J., Richardson, D. C., & Orgs, G. (2017). Joint action aesthetics. *Plos one, 12*(7), e0180101. https://doi.org/10.1371/journal.pone.0180101

von Zimmermann, J., Vicary, S., Sperling, M., Orgs, G., & Richardson, D. C. (2018). The choreography of group affiliation. *Topics in Cognitive Science, 10*(1), 80-94. https://doi.org/10.1111/tops.12320

Yeung, M. K. S., & Strogatz, S. H. (1999). Time delay in the Kuramoto model of coupled oscillators. *Physical Review Letters, 82*(3), 648–651.